\documentclass[a4paper,11pt]{article}
\pdfoutput=1 

\usepackage{jinstpub} 

\usepackage{lineno}
\usepackage{lipsum}
\usepackage{amsmath}
\usepackage{graphicx}
\usepackage{csquotes}
\usepackage{color}
\usepackage[dvipsnames]{xcolor}   
\usepackage{cleveref}
\usepackage[separate-uncertainty=true]{siunitx}  

\title{Investigating the slow component of the infrared scintillation time response in gaseous xenon}

\author[a,1]{R. Hammann,\note{Corresponding author.}}
\author[a]{K. B\"ose,}
\author[a]{L. H\"otzsch,}
\author[a]{F. J\"org,}
\author[a]{and T. Marrod\'an Undagoitia}

\affiliation[a]{Max-Planck-Institut f\"ur Kernphysik, Saupfercheckweg 1, 69117 Heidelberg,
Germany}

\emailAdd{robert.hammann@mpi-hd.mpg.de}

\newcommand{\lyHighP}{$(6347 \pm 22\mathrm{_{\,stat}} \pm 400\mathrm{_{\,syst}})\,\mathrm{ph/MeV}$}
\newcommand{\lyLowP}{$(4663 \pm 16\mathrm{_{\,stat}} \pm 300\mathrm{_{\,syst}})\,\mathrm{ph/MeV}$}

\abstract{
Xenon is the target material of choice in several rare event searches.
The use of infrared (IR) scintillation light, in addition to the commonly used vacuum ultraviolet (VUV) light, could increase the sensitivity of these experiments.
Understanding the IR scintillation response of xenon is essential in assessing the potential for improvement.
This study focuses on characterizing the time response and light yield (LY) of IR scintillation in gaseous xenon for alpha particles at atmospheric pressure and room temperature.
We have previously observed that the time response can be described by two components: one with a fast time constant of $\mathcal{O}(\mathrm{ns})$ and one with a slow time constant of $\mathcal{O}(\mathrm{\mu s})$.
This work presents new measurements that improve our understanding of the slow component.
The experimental setup was modified to allow for a measurement of the IR scintillation time response with a ten times longer time window of about \SI{3}{\micro s}, effectively mitigating the dominant systematic uncertainty of the LY measurement.
We find that the slow component at about \SI{1}{bar} pressure can be described by a single exponential function with a decay time of about \SI{850}{ns}.
The LY is found to be \lyHighP, consistent with our previous measurement.
In addition, a measurement with zero electric field along the alpha particle tracks was conducted to rule out the possibility that the slow component is dominated by light emission from drifting electrons or the recombination of electrons and ions.
}


\proceeding{Light Detection in Noble Elements (LIDINE 2023)\\
  20-22 September 2023\\
  Madrid, Spain}

\begin{document}
\newcommand{\comment}[1]{\textcolor{red}{#1}}
\newcommand{\finalize}[1]{\textcolor{cyan}{#1}}

\maketitle
\flushbottom

\section{Introduction}
\label{sec:setup}
Xenon in its gaseous and liquid state is successfully used as a detector target material for rare-event searches, such as the direct detection of dark matter \cite{Chepel:2012sj,Gonzalez-Diaz:2017gxo,XENON:2023wimp,LZ:2023wimp}.
For this purpose, the excellent scintillation properties of the noble gas in the vacuum ultraviolet (VUV) spectrum are exploited \cite{Jortner:1965s,Fujii_2015xx}.
In addition, xenon is also known to scintillate in the infrared (IR) regime \cite{Carungo:1998xx,Bressi:2000nim,Borghesani:2001xx}.
This raises interesting prospects for using the IR scintillation component to enhance background discrimination in experiments searching for small signals.
To explore this potential fully, a thorough understanding of the IR scintillation properties of xenon is required.

In a previous paper \cite{Piotter2023}, we have presented a measurement of the time response of the IR scintillation process and its light yield in xenon gas at atmospheric pressure and room temperature for alpha particles.
We found that the time response is composed of two components: a fast one with a time constant of $\mathcal{O}(\mathrm{ns})$ and a slow one with a time constant of $\mathcal{O}(\mathrm{\mu s})$, during which the majority of IR photons are emitted.
Due to a limited acquisition time window of about \SI{300}{ns}, we could not determine the function describing the slow component, which gave rise to the dominant systematic uncertainty in the light yield (LY) measurement.
In this paper, we present new measurements that improve our understanding of the slow component and remove the dominant systematic uncertainty in the LY measurement.

\begin{figure}[t]
    \centering
    \includegraphics[width=\textwidth]{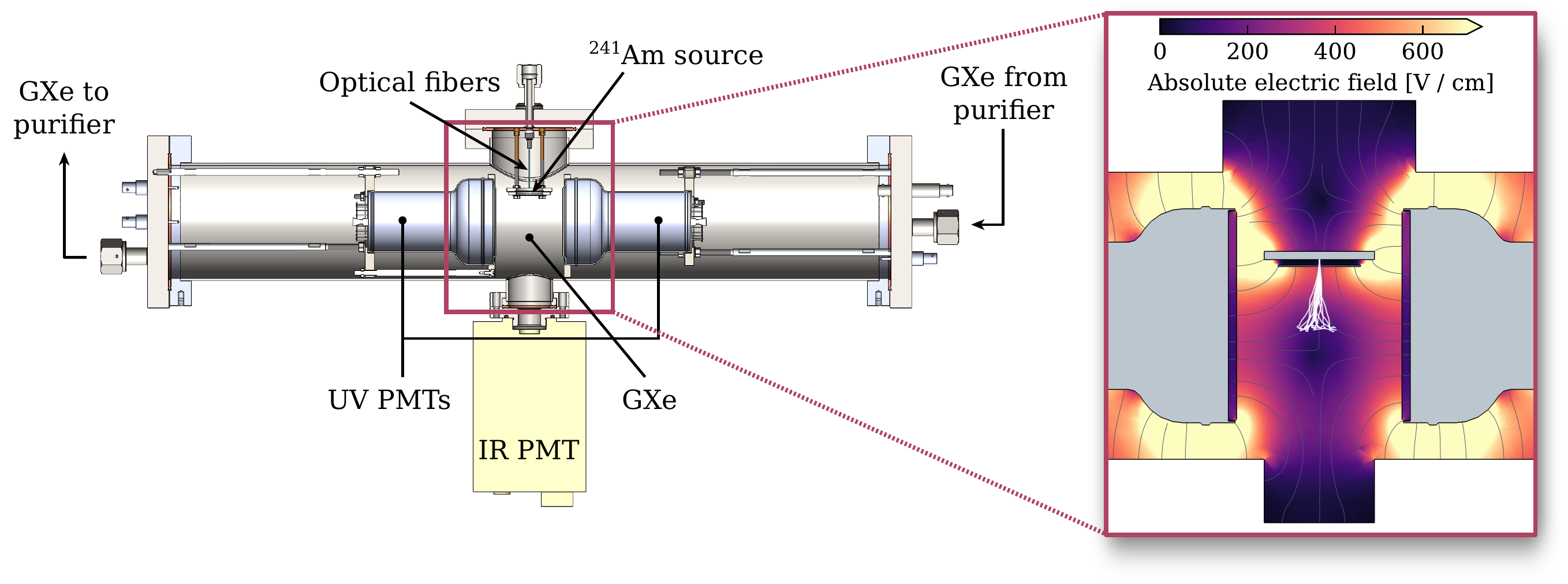}
    \caption{Schematic drawing of the experimental setup. The magnification shows a cut through a three-dimensional COMSOL simulation \cite{comsol:1998} of the electric field in the setup. A selection of alpha particle tracks simulated with SRIM \cite{Ziegler:2010} at 1 bar are overlaid in white. Figure adapted from \cite{Piotter2023}.}
    \label{fig:setup}
\end{figure}

The experimental setup is shown in \cref{fig:setup} and is described in detail in \cite{Piotter2023}.
It consists of a stainless steel tube filled with gaseous xenon and is instrumented with three photomultiplier tubes (PMTs).
Two of the PMTs are sensitive to UV light and are used to trigger the data acquisition (DAQ).
The third PMT is sensitive to IR light in the wavelength range from \SIrange{950}{1650}{nm} with a quantum efficiency of \SI{9.0\pm0.5}{\%}.
Importantly, it is sensitive to single IR photons with a time resolution of $\mathcal{O}(\SI{1}{ns})$.
The gas is irradiated with a collimated $^{241}$Am alpha source, featuring an energy of \SI{5.49}{MeV}.

All PMT signals are digitized using a CAEN V1743 ADC board.
In \cite{Piotter2023} we used the maximum sampling frequency of \SI{3.2}{GS/s}, which resulted in a maximum time window of \SI{320}{ns}.
To extend the time window in this work, we decreased the sampling frequency to \SI{1.6}{GS/s} and recorded six duplicates of the IR signal with relative digital time delays of \SI{500}{ns} between each duplicate.
This way, two consecutive time windows overlap by \SI{140}{ns}, and the total time window is extended to about \SI{3000}{ns}.

One unexplored aspect of our previous study was the influence of the non-vanishing electric field along the alpha particle track.
This inhomogeneous field, resulting from the negative high voltage applied to the UV PMT cathodes, varied between absolute values of zero and \SI{400}{V/cm} along the particle's trajectories (see magnification in \cref{fig:setup}).
To address this, we carried out a new measurement under conditions of zero electric field, which was achieved by operating the UV PMTs with the cathode at ground and anode at positive high voltage.
It should be noted, however, that the first measurements described below were made with the same inhomogeneous field as in \cite{Piotter2023}.

\section{Measurements and Results}
The measurements were performed with the experimental setup used in \cite{Piotter2023} with the modifications described above.
It was shown that the IR scintillation signal is strongly affected by the presence of impurities in the gas.
Thus, the gas is continuously purified by recirculating it through a zirconium hot-getter.
The recirculation was started at least 40 minutes prior to each measurement with a  recirculation mass flow approximately proportional to the pressure.
This ensured that the data was acquired in the regime of constant UV and IR light yields (see \cite{Piotter2023}).
A coincidence of two UV PMT signals triggers the digitization of all three PMT signals.
The time response of the IR scintillation process is obtained using the single photon counting technique, which is robust against signal distortion induced by electronics.
For this, the arrival times of individual IR photons are determined relative to the rising edge of the UV signal, and a histogram of $\mathcal{O}(10^6)$ IR photon arrival times is constructed (see \cite{Piotter2023} for details).

\subsection{Characterizing the slow scintillation component}
\label{sec:slow_component}
Here we present a new measurement of the IR time response with an acquisition time window of $\sim$\SI{3}{\micro s} at a pressure of \SI{1077.0\pm1.0}{mbar} and room temperature, which we compare to the measurement at \SI{1047.0\pm1.0}{mbar} with a $\sim$\SI{0.3}{\micro s} time window of our previous study.
We apply the cuts detailed in \cite{Piotter2023}: selecting alpha tracks centered between the UV PMTs via similar UV signal sizes, using a total UV area cut to eliminate pile-up and high energy loss events, and applying IR peak-specific cuts to reject PMT noise with low area and pulse width.
The IR pulse width cut is set at \SI{2.2}{ns} (compared to the previous value of \SI{1.85}{ns}) to better account for the observed distribution.

\begin{figure}[t]
    \centering
    \includegraphics[width=\textwidth]{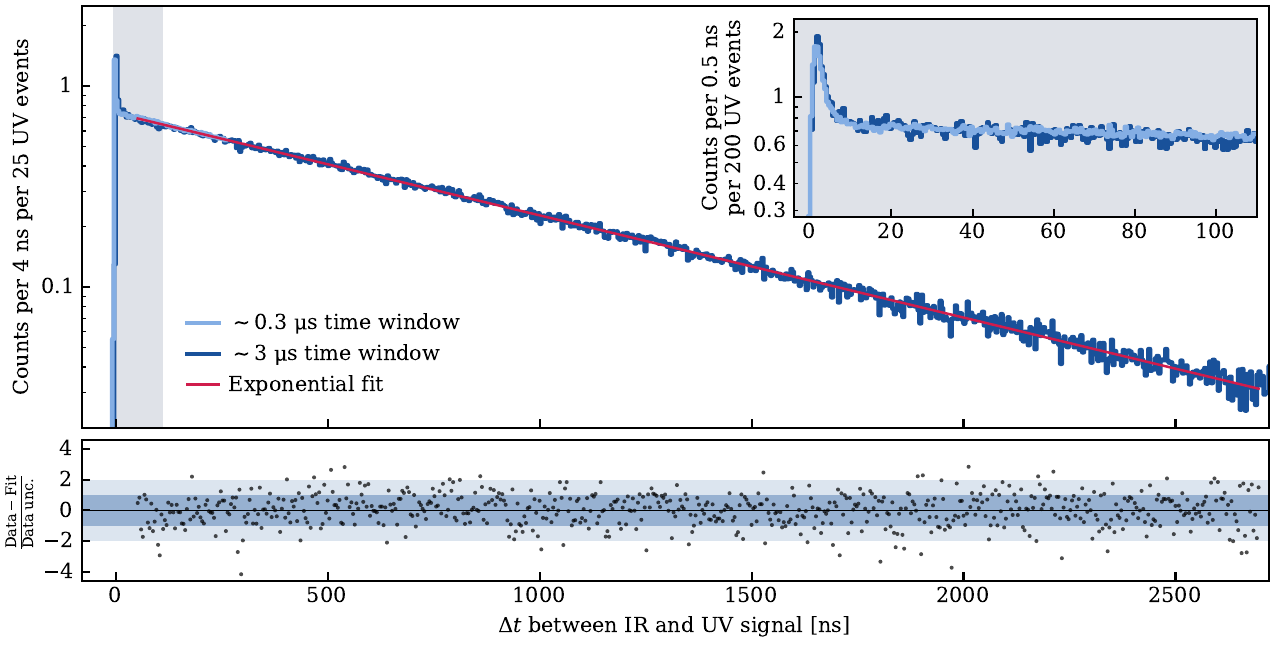}
    \caption{Time response of the IR scintillation process at \SI{1077.0\pm1.0}{mbar} with a time window of $\sim$\SI{3}{\micro s} (dark blue). For comparison, the previously obtained result with a time window of $\sim$\SI{0.3}{\micro s} is shown in light blue. The exponential fitted to the tail of the dark blue data is drawn in red. The bottom panel displays the residuals of the fit.}
    \label{fig:long_tw}
\end{figure}

The IR PMT baseline due to dark counts is estimated and subtracted from the final time response using a pre-trigger sample.
To address variations in peak detection efficiency among the digitizer channels, we use the information from the overlapping time windows.
The count overlap within the shared time window for the two channels is computed.
The histogram of the channel with the larger digital delay is then scaled by the ratio of the two counts.
The relative peak finding efficiency for the \SI{1.6}{GS/s} sampling frequency compared to the \SI{3.2}{GS/s} used in the previous study was determined to be $91.37\pm0.25\,\%$.
This value was obtained by comparing the number of events in the first $\sim$\SI{0.3}{\micro s} time window for the two sampling frequencies at about \SI{740}{mbar}, after verifying that the ratio is uniform within this time window.
For the \SI{3.2}{GS/s} sampling frequency, we assume an absolute peak finding efficiency close to $100\,\%$.
We assume that the lower efficiency of peak finding at the slower digitization rate is due to the fact that the number of voltage samples per SPE peak is reduced from about 10 to about 5, given their FWHM of about \SI{3}{ns}.
The time response obtained is shown in \cref{fig:long_tw}  along with the previous measurement that used a shorter time window.
The two measurements are in good agreement.

In our previous work, we employed three functions to fit the tail of the time response: a linear, an exponential, and a function proportional to $1/(1 + \Delta t/T_\mathrm{r})^2$ with the recombination time $T_\mathrm{r}$, motivated by the recombination process \cite{Kubota:1979xx}.
Due to the limited time window, none of these models could be ruled out in the previous measurement.
With the new measurement, we perform least-square fits with the same models in the time range between \SI{50}{ns} and \SI{2700}{ns}.
Through Pearson's $\chi^2$ goodness-of-fit (GOF) test we quantify that only the exponential model yields an acceptable fit with $\chi^2/ndf = 726/660 = 1.1$, corresponding to a p-value of $0.04$.
Even though this p-value is slightly below the conventional threshold of $0.05$, we consider the exponential model to be a good description of the tail.
The fit residuals are shown in the bottom panel of \cref{fig:long_tw}, which indicate a good agreement with the data, except for a few outliers below $-3\sigma$, which might be caused by digitizer effects.
We find a decay constant of  $(854.0 \pm 1.8\mathrm{_{\,stat}} ^{+26}_{-4}\mathrm{_{\,syst}})\,\mathrm{ns}$, where the systematic uncertainty is estimated by varying the fit range.
The same long time window measurement was performed at a pressure of \SI{743.0\pm1.0}{mbar} and room temperature, shown in \cref{fig:ly_pressure} (left).
Here, we find a decay constant of $(1064.8 \pm 2.8\mathrm{_{\,stat}} ^{+14}_{-0.9}\mathrm{_{\,syst}})\,\mathrm{ns}$ with a p-value of $0.13$ for the exponential fit.
The best-fit value agrees with our previous observation that the time constant increases with decreasing pressure.
The exclusion of the recombination model to characterize the tail suggests that the slow IR component is not dominated by scintillation after recombination.

\subsection{Light yield computation}
The light yield $Y$ is defined as the number of photons emitted per unit of deposited energy

\begin{equation}
    \label{eq:ly}
    Y = \frac{\mu}{E_\mathrm{\alpha} \cdot QE \cdot \epsilon} \,,
\end{equation}
where $\mu$ is the mean number of \textit{detected} photons per alpha event and $E_\mathrm{\alpha}$ is the energy deposited by the alpha particle behind the aperture.
The mean number of detected photons is obtained by integrating the time response, which takes into account the overlapping time windows and the small fraction of the signal that is outside of the total acquisition time window.
To obtain the number of \textit{emitted} photons we need to correct for the quantum efficiency of the PMT  $QE$, and the PMT's solid angle acceptance $\epsilon$.
The solid angle acceptance and the deposited energy for each gas pressure are estimated using a simplified Monte Carlo simulation, as described in \cite{Piotter2023}.
We are currently developing a more refined simulation that takes into account the topology of the alpha tracks, the quality cuts applied, and the optical properties of the materials.
The quantum efficiency is taken from the manufacturer's specification.

\begin{figure}[t]
    \centering
    \includegraphics[width=.49\textwidth]{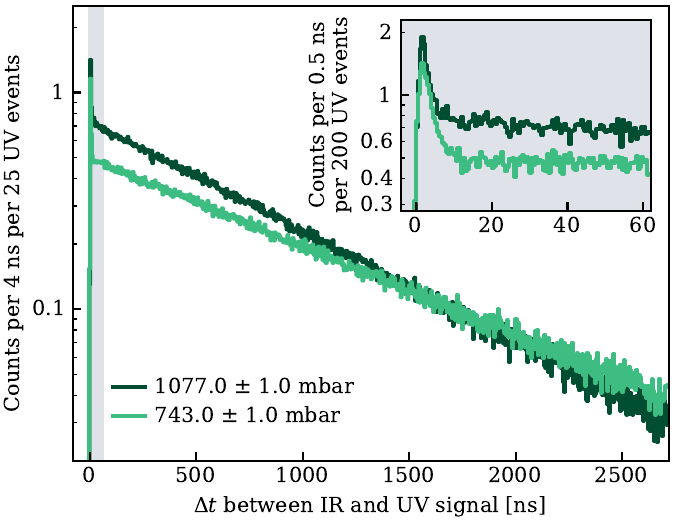}
    \includegraphics[width=.49\textwidth]{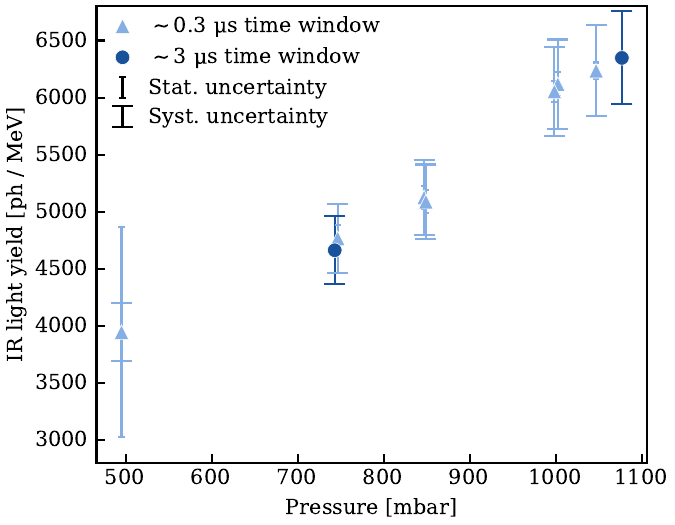}
    \caption{Left: Comparison of the IR scintillation time response for different pressures. Right: Pressure dependence of the IR light yield. The new measurements with long time window shown in dark blue follow the same trend as the previous measurements \cite{Piotter2023_data} shown in light blue. Note that the statistical uncertainty error bars of the new data points are not visible as they are smaller than the size of the marker.}
    \label{fig:ly_pressure}
\end{figure}

For the measurement at \SI{1077.0\pm1.0}{mbar} and room temperature, we obtain a light yield of \lyHighP.
At \SI{743.0\pm1.0}{mbar}, we find \lyLowP.
Both measurements are plotted in \cref{fig:ly_pressure} (right) together with the previous measurements, which show good agreement.
In particular, they follow the same increase of the light yield with pressure.
To estimate the systematic uncertainty, we took into account uncertainties due to the PMT gain, the deposited energy behind the aperture $E_\alpha$, and the solid angle acceptance $\epsilon$, as well as the variation of the $QE$ over the sensitive wavelength band, which is the dominant contribution.
The relative systematic uncertainty is about $6\,\%$.
Notably, this contribution is about an order of magnitude smaller than the previously dominant systematic uncertainty on the fit function.

\subsection{Zero-field measurement}
In our previous study and the measurements described above, the setup exhibited a non-zero electric field along the alpha particle tracks.
To explore potential sources of the slow component, including light emission from drifting electrons via neutral bremsstrahlung \cite{henriques2022neutral} and scintillation after recombination, measurements were conducted with zero electric field along the alpha particle tracks.

Due to problems with HV breakdowns, we were only able to take measurements with one of the UV PMTs at a reduced high voltage and the other one turned off.
The UV signals were still sufficiently large to determine the rising edge of the signal clearly, which is required for timing measurements.
To adapt to the single UV channel, the coincidence level was set to one, and adjustments were made to the data quality cuts.

\begin{figure}[t]
    \centering
    \includegraphics[width=\textwidth]{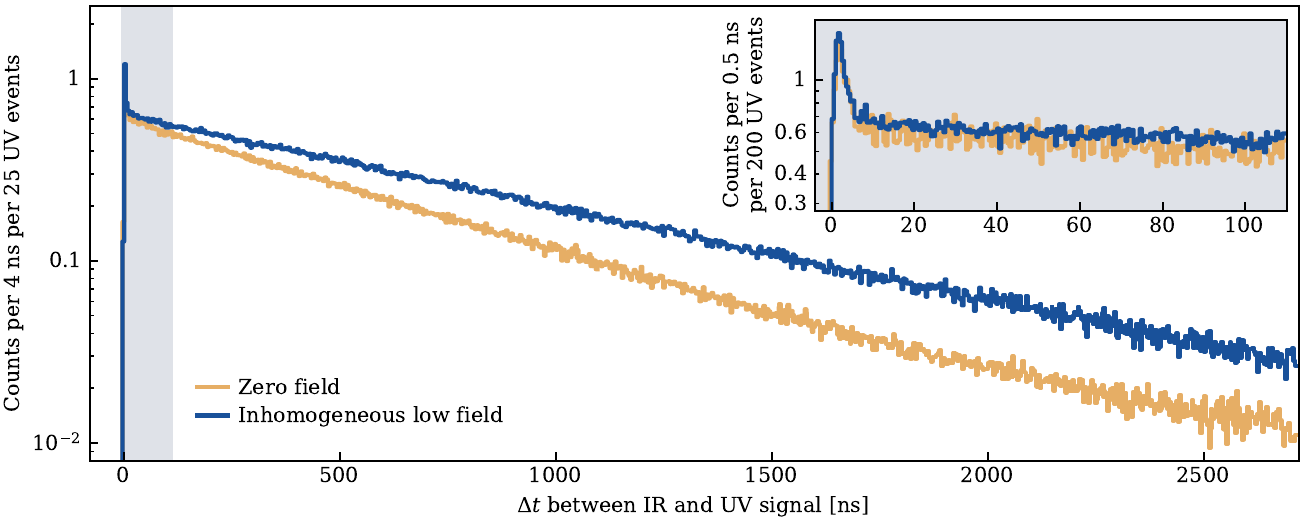}
    \caption{Absolute time response of the IR scintillation process for the zero-field measurement compared to the measurement with non-zero field (\SIrange{0}{400}{V/cm} along the alpha particle tracks).}
    \label{fig:zero_field}
\end{figure}

The time response of the IR scintillation process for the zero-field measurement at \SI{1057.0\pm1.0}{mbar} is shown in \cref{fig:zero_field}.
The fast component remains consistent with the measurement at non-zero field (\SI{1077.0\pm1.0}{mbar}), while the slow component persists with a comparable amplitude but a faster decay constant of approximately \SI{600}{ns}.
Since the setup was opened between the two measurements to replace the PMT bases, we are currently investigating whether the different decay constant for the zero-field measurement might originate from a slightly higher level of impurities after the modifications.
However, the persistence of the slow component, without an increase in amplitude allows us to draw conclusions on its origin.
It rules out the possibility of the slow component being dominated by light emission from drifting electrons, as it would have vanished under zero-field conditions.
Additionally, we can conclude that scintillation after recombination can not be a dominant constituent of the slow component, as its amplitude would have increased in the absence of an electric field.

\section{Discussion and Outlook}
\label{sec:discussion}
In agreement with our previous measurement \cite{Piotter2023}, we have found that the IR scintillation process is composed of two components: a fast one with a decay time of a few ns and a slow one with a decay time of $\mathcal{O}(\mu \mathrm{s})$, which dominates the light yield (fraction of the fast component $<$ \SI{1}{\%}).
With a ten times longer time window of about \SI{3}{\micro s} compared to our previous measurement, we can conclude that the slow component is well described by a single exponential function with a decay time of about \SI{850}{ns} at atmospheric pressure.
We determine the light yield at \SI{1077.0\pm1.0}{mbar} and room temperature to be \lyHighP.
This is about a factor three lower than the light yield of \SI{20000}{ph/MeV} measured by \cite{Belogurov:2000}, which is the only other measurement of the IR light yield in gaseous xenon.
Note that this measurement was performed at a pressure of \SI{2}{bar} and with a considerably different experimental setup featuring an InGaAs photodiode.
Our reported IR light yield is approximately five times lower than the UV light yield of approximately \SI{25000}{ph/MeV} \cite{Henriques:2023UV_LY,Leardini:2021qnf}.
Another measurement at \SI{743.0\pm1.0}{mbar} confirms the previously observed increase of the light yield with pressure.
The origin of the pressure dependence is currently not understood.

Through a measurement with zero electric field along the alpha particle's trajectory, we have excluded that the slow component is dominated by light emission due to drifting electrons as well as scintillation after recombination.
The latter is further supported by the inadequacy of the model derived for the recombination process in describing the slow component.
In addition, with electron drift velocities below \SI{2}{mm\per\micro\second} expected for the electric fields in our original setup \cite{Dias1993}, we roughly estimate that the electron drift time exceeded \SI{15}{\micro \second}. This is about an order of magnitude slower than the observed time constant.
Nevertheless, despite the progress made in this work, the origin of the slow component remains unclear.


To improve our understanding of the scintillation process in gaseous xenon, and in particular of the slow IR component, we intend to conduct wavelength-band resolved measurements of the time response similar to \cite{Leardini:2021qnf}.
Additionally, we aim to carry out measurements at higher pressures to further investigate the pressure dependence of the LY.
Eventually, we plan to perform measurements in liquid xenon to investigate the potential application of the IR scintillation component to improve future double-phase liquid xenon detectors \cite{Aalbers:2016jon,Aalbers:2022dzr}.

\acknowledgments

We would like to thank the organizers of LIDINE 2023 for hosting this great conference and giving us the opportunity to present our work.
Our sincere thanks go to the numerous attendees for the engaging discussions throughout the conference.
In particular, we would like to thank Diego Gonz\'alez D\'iaz and Armando Francesco Borghesani for extensive and insightful conversations regarding the IR scintillation process in xenon.
We also thank Vicente Pesudo Fortes, Douglas Fields, and Austin de St. Croix for their constructive discussions.
Additionally, we want thank Dominick Cichon for his unwavering support of the project.
We acknowledge the support of the Max Planck Society.

\clearpage
\bibliographystyle{JHEP}
\bibliography{references}   
\end{document}